\documentclass[prb,notitlepage,twocolumn]{revtex4}

\usepackage{amsmath}
\usepackage{amssymb}
\usepackage{bm}
\usepackage{graphicx}
\usepackage{times}
\usepackage{color}

\usepackage{wrapfig}
\begin{document}

\title{Functional  meta-optics and nanophotonics govern by Mie resonances}

\author{Sergey Kruk$^{1}$}
\author{Yuri Kivshar$^{1,2}$}

\affiliation{$^1$Nonlinear Physics Center, Australian National University,~Canberra ACT 2601,~Australia\\
$^2$ITMO University, St. Petersburg 197101, Russia}

\begin{abstract}
Scattering of electromagnetic waves by subwavelength objects is accompanied by the excitation of electric and magnetic Mie resonances,
that may modify substantially the scattering intensity and radiation pattern. Scattered fields can be decomposed into electric and magnetic multipoles, and the magnetic multipoles define magnetic response of structured materials underpinning the new field of {\em all-dielectric resonant meta-optics}.
Here we review the recent developments in meta-optics and nanophotonics, and demonstrate that the Mie resonances can play a crucial role offering novel ways for the enhancement of many optical effects near magnetic and electric multipolar resonances, as well as driving a variety of interference phenomena which govern recently discovered novel effects in nanophotonics. We further discuss the frontiers of all-dielectric meta-optics for flexible and advanced control of light with full phase and amplitude engineering, including nonlinear nanophotonics, anapole nanolasers, quantum tomography, and topological photonics.\\

\color{blue} \bf{Accepted to ACS Photonics}
\end{abstract}

\maketitle


It is well known that natural materials do not demonstrate any magnetic properties  at optical frequencies, this is because a direct action of the optical magnetic field on matter is much weaker than electric ones.  Nevertheless, the manifestation of {\em "optical magnetism"}  is found in specifically designed artificial subwavelength structures that allow strong magnetic response, even when such structures are made of non-magnetic materials. This progress became possible due to the fascinating field of {\em electromagnetic metamaterials} that describes
optical structures composed of subwavelength elements, often called "meta-atoms", and specific localized fields these structures may support, creating a platform for {\em metadevices}~\cite{zheludev2012metamaterials}.

It has been established that the use of artificial "meta-atoms" would allow to engineer magnetic permeability
$\mu$ and magnetic response by achieving strong resonances in structured systems made of non-magnetic materials.
Although real magnetism in its conventional sense is not available at high optical frequencies, it is possible to engineer
the spatial dispersion and nonlocal electric effects in such a way to induce a strong magnetic dipole moment even though the involved materials do not possess microscopic magnetization.

Classical "meta-atom" is a metallic split-ring resonator where electrons oscillate back and forth creating an effective loop of current, and thus, an efficient magnetic response. The concept of split-ring resonators was first introduced at microwaves to realize artificial magnetic inclusions with subwavelength footprint, and then it was translated to the optical spectral range exploiting the plasmonic features of metallic nanoparticles~\cite{review}. By now, this concept was realized in many non-magnetic {\em plasmonic structures} ranging from nanobars~\cite{cai_2007} and nanoparticle complexes often called "oligomers" ~\cite{alu_2008,sun_2016} to split-ring-based structures~\cite{kante_2012,qin_2013} and more complicated structures
associated with hyperbolic-type magnetic response of fishnet metamaterials~\cite{kruk_2016}.

Historically, the subwavelength localization of light in nanophotonics was associated with free electrons and electromagnetic
waves at metallic interfaces studied by {\em plasmonics}. However, the recent developments of the physics of high-index dielectric
nanoparticles~\cite{science_review} suggests an alternative mechanism of light localization via low-order dipole and multipole
Mie resonances that may generate magnetic response via the displacement current contribution~\cite{merlin_2009}.

The study of resonant dielectric nanostructures has been established as a new research direction in modern nanoscale optics and metamaterial-inspired nanophotonics~\cite{science_review,isabelle_review}. Because of their unique optically induced electric and magnetic resonances, high-index nanophotonic structures are expected to complement or even replace different plasmonic components in a range of potential applications. We emphasize that the physics of all-dielectric resonant nanophotonics governed by Mie resonances in high-index dielectric nanoparticles can be broadly characterized by two major phenomena:

\begin{itemize}

\item

{\bf localization of electric and magnetic fields at the nanoscale} due to Mie resonances, which enhance
nonlinear effects, such as Raman scattering, harmonic generation, etc;

\item

{\bf engineering of multimodal interference} for driving many novel effects such as unidirectional scattering,
optical antiferromagnetism, bound states in the continuum, nonradiating optical anapoles, etc.

\end{itemize}

The resonant behavior of light in high-index dielectric nanoparticles allows to reproduce many subwavelength effects demonstrated in plasmonics due to the electric field localization, but without much losses and energy dissipation into heat. In addition, the coexistence of strong electric and magnetic multipolar resonances, their interference, and resonant enhancement of the magnetic field in dielectric nanoparticles bring many novel functionalities to simple geometries largely unexplored in plasmonic structures, especially in the nonlinear regime and metadevice applications.

This review paper aims to provide a broad view on the rapidly developing field of all-dielectric resonant
meta-optics and nanophotonics, uncovering a great potential of optically-induced electric and magnetic Mie resonances
for the design of a new generation of optical
 metadevices. We also discuss the emerging opportunities of meta-optics
driven by Mie resonances for novel physics of multimodal interference, and envisage the rapid progress for achieving
flexible control of light at the nanoscale with smart engineering of phase and amplitude for functional nanophotonics.

\section*{Enhanced optical effects}

\begin{figure}[t]
\includegraphics[width=0.46\textwidth]{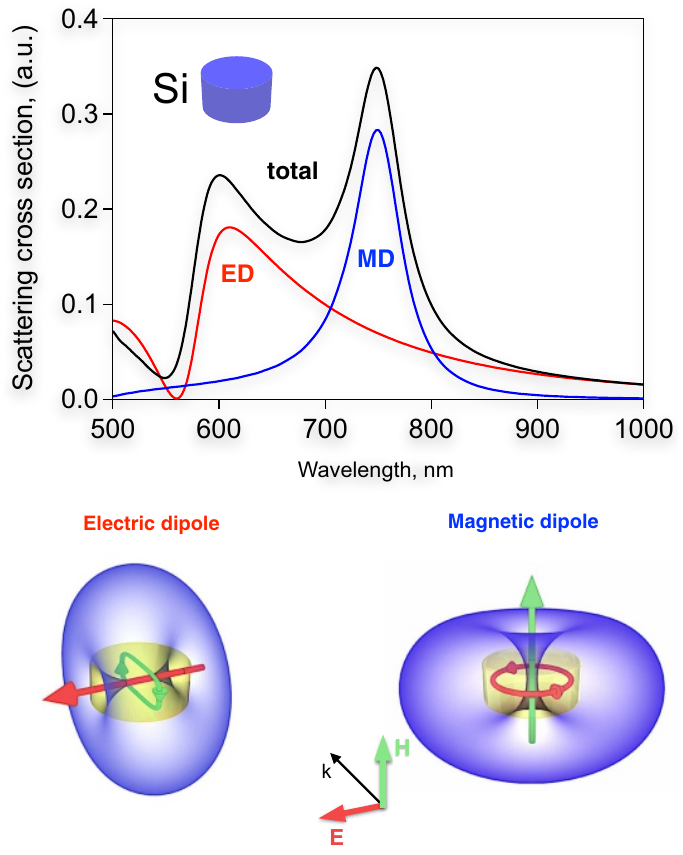}
\caption{\textbf{Fundamentals of Mie resonances.} Scattering efficiency of a silicon nanodisk with 200 nm radius and 260 nm height (see the insert) with the resonant contributions from magnetic dipole (MD) and electric dipole (ED) Mie resonances, with the schematics of their radiation patterns shown below.
}
\label{fig1}
\end{figure}

The physics of interaction between light and subwavelength nanoparticles dates back to works of Lord Rayleigh and Gustav Mie. Theories of Rayleigh scattering and Mie scattering described successfully such phenomena as changes of sky hue, colors of colloidal solutions and other. Today, with the advancements of nanotechnology, we employ the scattering theories for reverse engineering of subwavelength light-matter interactions: we design and fabricate nanoparticles with scattering properties at will~\cite{science_review}.
 To illustrate the properties of light scattering by nanoparticles, we consider a small nanodisk illuminated by a plane wave. The scattering of such a nanoparticle demonstrates strong resonances, as shown in Fig.~\ref{fig1}.

In a sharp contrast with plasmonics, for high-index dielectric nanoparticles of simple geometries, we can observe both electric and magnetic dipole resonances of comparable strengths. A strong {\em magnetic dipole (MD) resonance} appears due to a coupling of incoming light to the circular displacement currents of the electric field, owing to the field penetration and phase retardation inside the particle. This occurs when the wavelength inside the particle becomes comparable to its spatial dimension, $2R \approx \lambda/n$, where $n$ is the refractive index of particle’s material, $R$ is the nanoparticle radius, and $\lambda$ is the wavelength of light.  Such a {\em geometric resonance} suggests that a subwavelength nanoparticle ($R < \lambda$) should have a large refractive index in order to have the resonances in visible and IR spectral regions.  In addition, the excited MD response may become comparable (or even be stronger) with the electric dipole (ED) response providing a major contribution to the scattering efficiency.

In the visible and near-IR spectral ranges, the large permittivity is known to occur for semiconductors such as Si, Ge, and AlGaAs. In the neighbouring mid-IR range, which is also of great interest to nanophotonics, narrow-band semiconductors (Te and PbTe) and polar crystals such as SiC can be implemented for all-dielectric resonant metadevices driven by Mie resonances.  A search for better materials and fabrication techniques for high-index nanophotonics is an active area of research~\cite{baranov_optica}.

The first experimental observations of strong magnetic response in visible and IR have been reported for spherical Si nanoparticles~\cite{exp1,exp2}.  This opened a way towards the study of many interesting phenomena with high-index dielectric structures employed as building blocks for metadevices driven by inspiration from century-old studies of light scattering.

\begin{figure*}[t]
\includegraphics[width=0.9\textwidth]{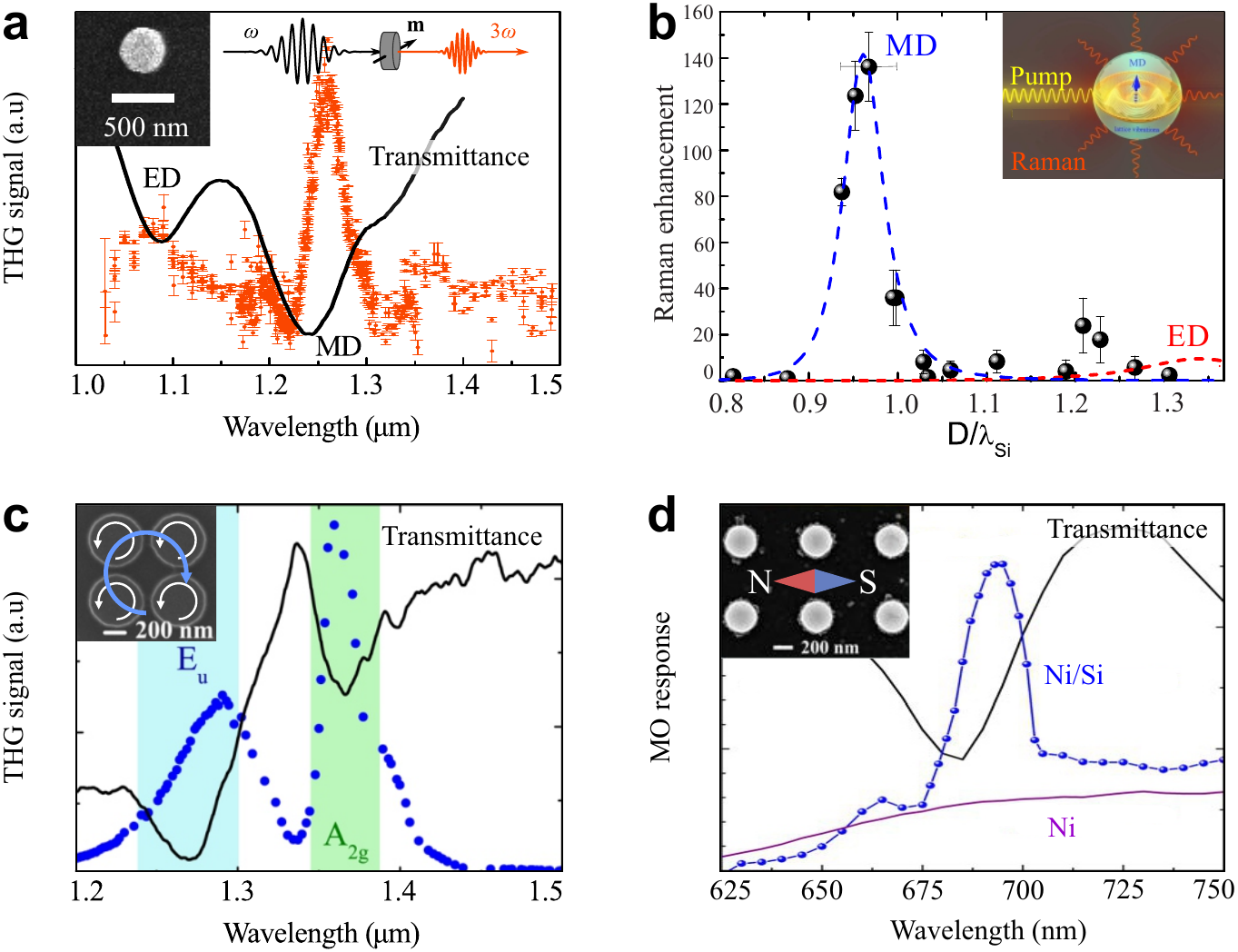}
\caption{\textbf{Enhancement of optical effects due to the magnetic dipole Mie resonance.}
(a) Nonlinear THG from Si nanodisks~\cite{melik}, (b) enhanced Raman scattering from
a spherical Si nanoparticle~\cite{krasnok}, (c) magnetic Fano resonance in nanoparticle quadrumers~\cite{shorokhov},
and (d) multifold enhancement of the magneto-optics response when a Si nanoparticle is covered by a Ni film~\cite{barsukova}.
Inserts show the geometries of nanostructures.
}
\label{fig2}
\end{figure*}

One of the main concepts underpinning the field of plasmonics is the large enhancement of the electric field in hot-spots
of metallic nanoparticles. However, plasmonic resonances are accompanied by large dissipative losses. Two strategies to employ
high-index dielectrics instead of metals are the creation of hot-spots for electric and magnetic fields~\cite{samusev,polman},
and the use of large concentration of the incident electromagnetic field inside the dielectric particles~\cite{trib_pra,trib_sr}.
A giant enhancement of electromagnetic fields has recently been detected experimentally at microwaves,
with the intensity of the magnetic field more than 300 times larger than that of an incident wave, provided the Mie
resonant condition is satisfied~\cite{trib_sr}.  These results suggest  a substantial enhancement of many optical effects with nanoparticles
by \emph{at least two orders of magnitude}, and some selected examples are presented in Figs.~\ref{fig2}(a-d).

Figure~\ref{fig2}(a) summarizes the results of the third-harmonic generation (THG) from Si nanodisks at the ED and MD resonances~\cite{melik}. The enhanced upconversion efficiency at the MD resonance of a nanodisk is observed owing to the strong field confinement, whereas considerably lower average local field density and overall THG is detected at the ED resonance.
Similarly, the 140-fold enhanced Raman signal has been observed experimentally for individual spherical Si nanoparticles at the
magnetic dipole resonance~\cite{krasnok}, see Fig.~\ref{fig2}(b).

Thus, strong Mie-type MD resonances in all-dielectric nanostructures provide novel opportunities for enhancing many
optical effects at the nanoscale. One recent example is the substantial enhancement of the THG signal from
Si nanoparticle quadrumers associated with {\em magnetic Fano resonance}~\cite{shorokhov}. This effect is accompanied by
nontrivial wavelength and angular dependencies of the generated harmonic signal featuring a multifold enhancement
of the nonlinear response in this oligomeric nanoscale system [see Fig.~\ref{fig2}(c)].

Last but not least, we mention a very recent experimental demonstration~\cite{barsukova} of the enhanced magneto-optical effects in subwavelength
a-Si nanodisks covered with a thin nickel (Ni) film [shown in Fig.~\ref{fig2}(d)], and achieved the five-fold enhancement
of the magneto-optical response of hybrid magnetophotonic metasurfaces, in comparison with a thin Ni film deposited
on a Si substrate.

\section*{Multimodal and multipolar interference}

To understand the major characteristics of the light scattering at the nanoscale, one usually employs the multipole decomposition of electromagnetic fields. In this way, the Mie scattering is characterized by partial intensities and radiation patterns
of dominant excited multipole modes, not only the electric and magnetic dipoles but also higher-order multipolar modes
and the modes with the toroidal geometry. For metallic nanoparticles, the electric dipole mode usually dominates the Mie scattering.
In contrast, incorporating the optical magnetic response for dielectric particles provides an extra degree of freedom for efficient
light control, through interference of electric and magnetic dipoles and multipolar modes.

\begin{figure*}[t]
\includegraphics[width=1.0\textwidth]{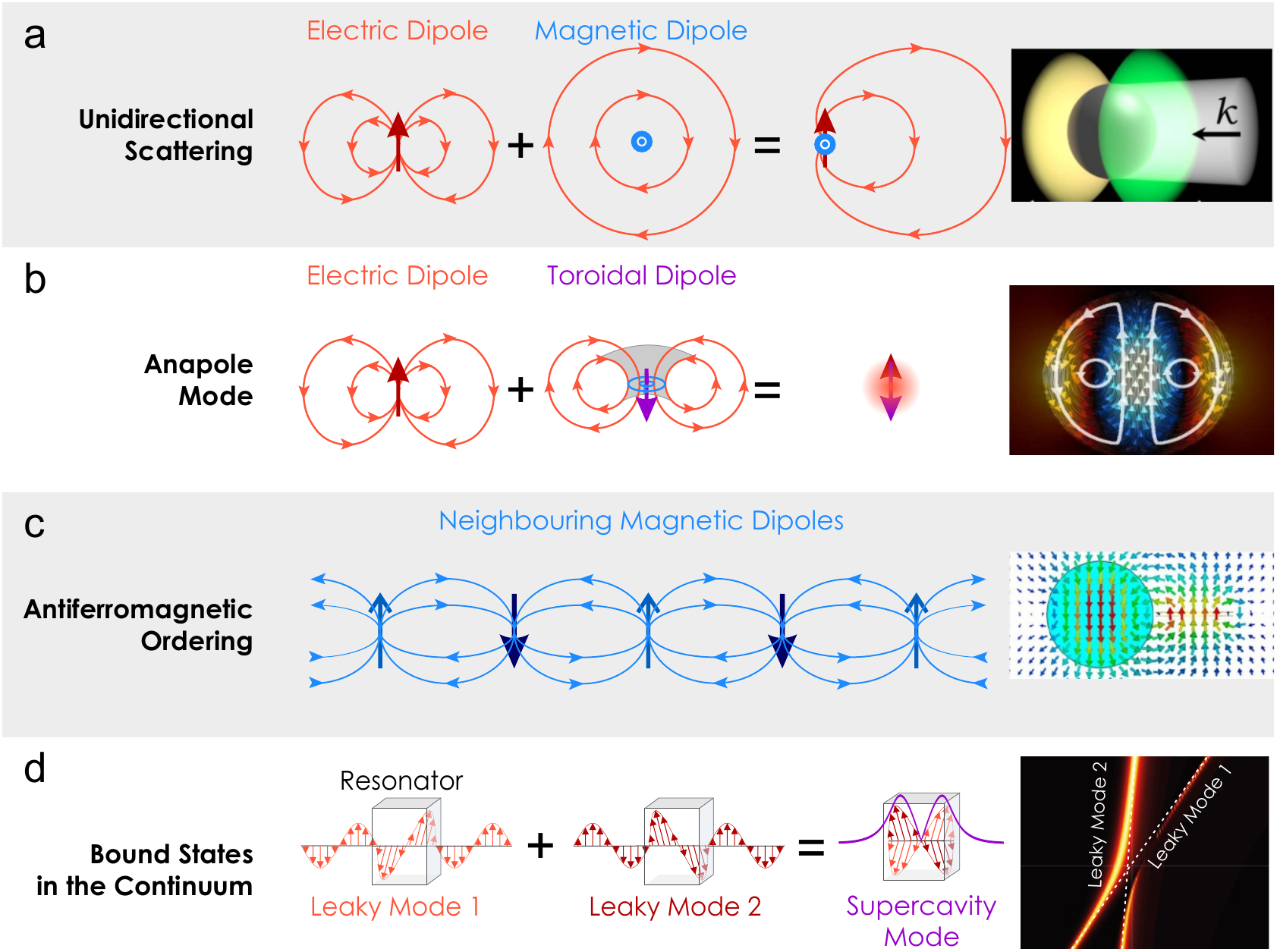}
\caption{\textbf{Multimodal and multipolar interference.} Examples of several major interference effects in all-dielectric resonant meta-optics
(a) Kerker's effect and unidirectional scattering~\cite{uni_ncomm}; (b) near-field confinement
of light in the anapole mode~\cite{anapole}; (c) antiferromagnetic ordering of magnetic dipole moments~\cite{afm};
and (d) strong mode coupling resulting in bound states in the continuum~\cite{bic}.}
\label{fig3}
\end{figure*}

In the simplest case, when multipole modes are of the same type and they are produced by different elements, the interference
physics of dielectric nanoparticles is associated with {\em Fano resonances}, and it resembles its plasmonic analogues~\cite{fano_review}.
Nanoparticle structures demonstrate sharp Fano resonances with characteristic asymmetric scattering due to interference between
non-radiative and radiative modes. Importantly, MD resonances of individual dielectric particles
play a crucial role in the appearance of the Fano resonances~\cite{fano_nl,fano_limonov}.

Interference between the electric and magnetic dipole modes can lead to {\em unidirectional scattering}, characterized by the enhancement of forward scattering and suppression of backward scattering, or vise versa. Strong directional scattering of light can be realized due to interference of magnetic and electric dipole responses excited simultaneously with comparable strength~\cite{uni_ncomm,novotny}, as shown in Fig.~\ref{fig3}(a). This is conceptually similar to the so-called {\em Kerker condition}~\cite{kerker} for the zero backward scattering, derived by M. Kerker {\em et al.} for a hypothetical magnetic particle having similar electric and magnetic properties.

Interference of electric and magnetic dipole modes offers only one example of a larger class of {\em multipolar interference effects} observable with dielectric particles. The approach extends further to higher orders as well as higher number of multipoles. One remarkable example is a control over a unidirectional scattering in a direction perpendicular to the direction of incident wave propagation by the interference of in-plane electric dipole and electric quadrupole together with an out-of plane magnetic dipole~\cite{Li_NanoLett_2016}.

Perhaps most unusual example of multipolar interference is the interference of the electric dipole and toroidal mode [see Fig.~\ref{fig3}(b)], which can suppress far-field radiation to create a virtually nonradiating source called {\em an optical anapole}~\cite{anapole}. Isolated excitation of toroidal modes requires careful engineering of the nanoparticle structure to match the special corresponding near-field current distributions; still, even simple homogeneous dielectric particles with Mie resonances can support modes with toroidal symmetry. Recently, optical anapoles have been experimentally demonstrated in isolated Si and Ge nanoparticles~\cite{anapole,anapole2}.

Interference between magnetic dipole modes of composite nanostructures can result in a new type of
optical magnetism resembling the staggered structure of spins in antiferromagnetic (AFM) ordered materials~\cite{afm,afm2}.
This effect was first predicted for a hybrid meta-atom formed by a metallic split-ring resonator and
high-index dielectric particle, both supporting optically-induced magnetic dipole
resonances of different origin~\cite{afm}. Such a ‘metamolecule’ is characterized by two types of interacting
magnetic dipole moments with the distance-dependent magnetization resembling the spin exchange
interaction in magnetic materials [see Fig.~\ref{fig3}(c)]. Recently, this prediction was demonstrated in experiment
by directly mapping the structure of the electromagnetic fields, with the observation of strong coupling between
the optically-induced  magnetic moments accompanied by a flip of the magnetisation orientation
in each metamolecule creating an AFM lattice of staggered optically-induced magnetic
moments~\cite{afm2}. Such AFM states can also be found in all-dielectric structures, but their importance in shaping the radiation field is still to be understood.

Finally, we mention that high-index dielectric nanoparticles have been suggested recently for achieving large values
of $Q$-factors of subwavelength nanoscale resonators in the regime of {\em bound states in the continuum}~\cite{bic}.
This occurs through strong coupling of modes with avoided crossing [see Fig.~\ref{fig3}(d)] and Fano resonances in dielectric
finite-length nanorods resulting in high-$Q$ nanophotonic \emph{supercavity}~\cite{nphot_nv}. For an individual Si nanoresonator, this
approach allows engineering supercavity modes to achieve $Q=200$, that is of paramount interest for nonlinear optics and
quantum nanophotonics, thus opening up new horizons for active and passive nanoscale metadevices.

Unidirectional scattering, anapole modes, optical antiferromagnetism, and bound states in the continuum are among the key fundamental concepts that will drive the progress in novel nanoscale photonic technologies in near future. We further review other emerging fields of meta-optics arising from
those concepts, but notice that many other ideas are expected to be developed in the coming years.

\section*{Planar meta-optics}

Unidirectional scattering of dielectric nanoparticles brings a new paradigm to the field of meta-optics. By arranging the forward-scattering nanoparticles in a planar layout, we can create a thin conformal layer that is resonant, yet transparent. Such two-dimensional arrangements of resonant elements are termed {\em metasurfaces}. In this section, we review the basic strategies to engineer the electromagnetic
space for an efficient control of light with highly transparent metasurfaces.

\begin{figure*}
\includegraphics[width=0.99\textwidth]{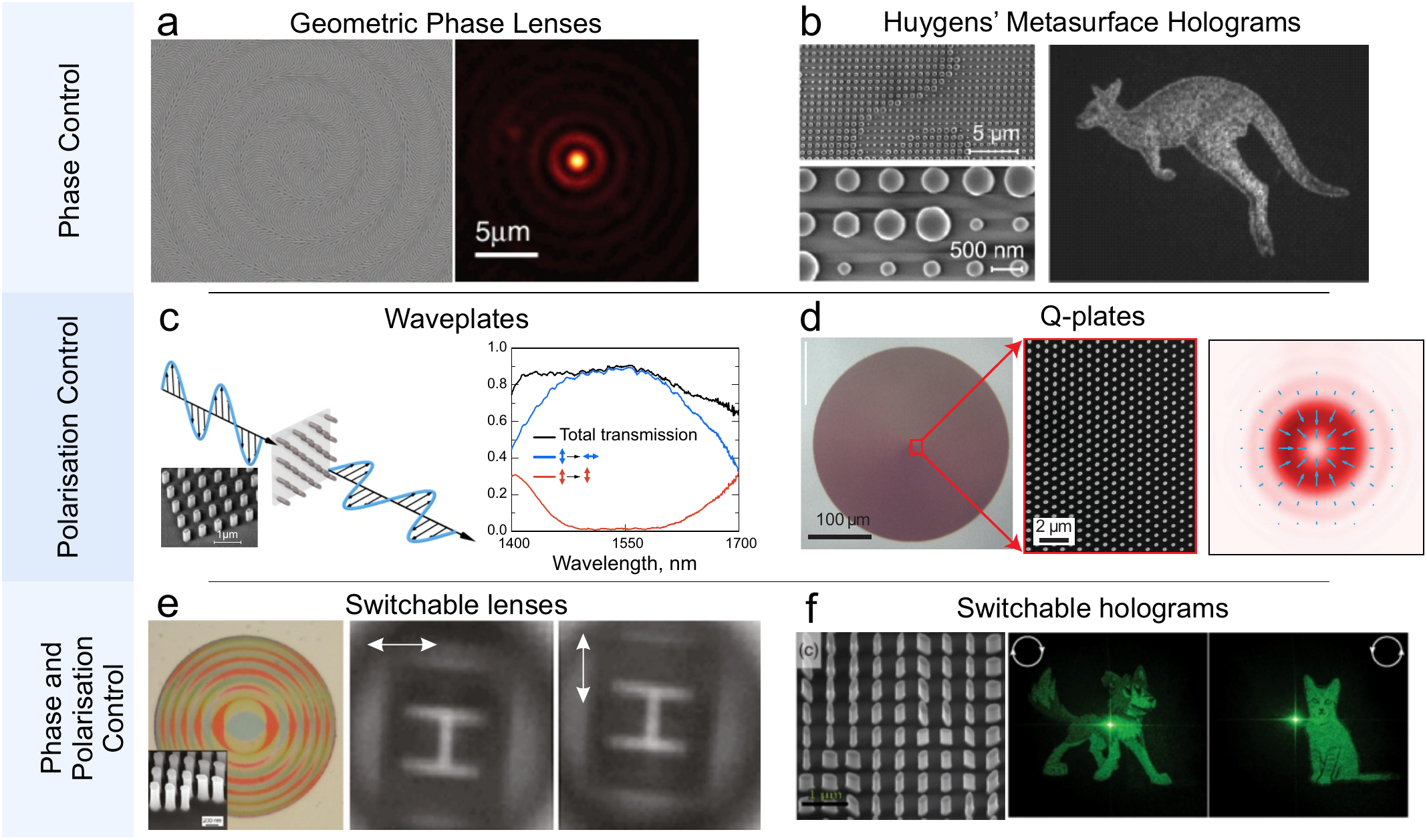}
\caption{\textbf{Examples of planar metadevices and their functionalities.} (a,b) Phase control with resonant metasurfaces by using (a) geometric phase~\cite{lin2014dielectric} and (b) generalised Huygens' principle~\cite{wang2016grayscale}. (c,d) Polarisation control with form-birefringent metasurfaces. (c) Periodic array of anisotropic nanoresonators operating as a half-wave plate~\cite{kruk2016invited}. (d) Metasurface $q$-plate generating radially polarised cylindrical vector beams~\cite{arbabi2015dielectric}. (e,f) Simultaneous control of phase and polarisation for polarisation switchable (e) meta-lenses~\cite{schonbrun2011reconfigurable} and (f) meta-holograms~\cite{mueller2017metasurface}.}
\label{fig:Part3Fig1}
\end{figure*}

{\bf Huygens' metasurfaces.} An unidirectional scatterer that satisfies Kerker's condition represents an object described back to 1690: {\em a Huygens' source}~\cite{Huygens1690}. A well-known Huygens' principle states that each point on a wavefront acts as a secondary source of outgoing waves. The principle implies that sources do not radiate backward, however in its original formulation this principle does not specify the structure of the sources that would satisfy this requirement. More recently, it was suggested that a Huygens' source can be realised as electrically small antenna that is a superposition of crossed electric and magnetic dipoles.~\cite{Krasnok_JETP,geffrin2012magnetic} Thus, a dielectric nanoparticle that supports both electric and magnetic dipole modes does fulfil these requirements. An array of such nanoparticles creates a Huygens' metasurface that is not only transparent to incident light but, being resonant, allows for the phase accumulation in the entire range 0~$\div$~2$\pi$. Indeed, a single resonance, at either electric or magnetic dipole, gives up to the $\pi$ phase accumulation depending on its spectral position. When two resonances overlap, the total phase accumulation reaches 2$\pi$~\cite{decker2015high}. Importantly, by creating such a Huygens' metasurface from different elements tuned to different resonant frequency, one can create subwavelength phase gradients in the region 0~$\div$~2$\pi$ thus allowing for a full wavefront control of light by a subwavelength planar device.

{\bf Generalised Huygens' conditions.} Huygens' metasurfaces prove to be a powerful concept with one limitation: a narrow spectral bandwidth around a single operational wavelength. Generalised Huygens' principle~\cite{kruk2016invited} overcomes this limitation. We notice here that the regime of the unidirectional scattering is not limited to dipolar resonances. Exemplarily, an array of superimposed electric and magnetic quadrupoles would interfere similarly to an array of electric and magnetic dipoles. More generally, a unidirectional scattering can be achieved by overlapping several multipoles of both even and odd symmetries that can interfere destructively in backward direction and constructively in forward direction. This concept allows to achieve an extended spectral range of operation~\cite{kruk2016invited}.

{\bf Form birefringence.} In addition to the phase control, dielectric nanoparticles allow to control polarisation of light via \emph{form birefringence}. Conventionally, birefringent response of optical materials originates from anisotropy of their crystalline lattices (such as that of calcite). However, anisotropy at the material level is not a fundamental requirement for birefringence. A dielectric nanoparticle made of an isotropic material can alter polarisation of light due to its anisotropic form as its multipolar scattering spectrum becomes polarisation-dependent. Similar to optical magnetism, the form birefringence of nanoscale objects rises from their geometry, not material properties. An array of anisotropic nanoparticles can change polarisation of light in a desired way~\cite{kruk2016invited,arbabi2015dielectric,mueller2017metasurface}. By combining nanoparticles of different shapes, one can create high-resolution subwavelength polarisation gradients, which is difficult to achieve with conventional anisotropic crystals
or even with liquid-crystal light modulators.

{\bf Geometric phase.} An ability to engineer polarisation of incident light opens another opportunity for the wavefront control via the geometric phase~\cite{pancharatnam1956generalized,berry1984quantal,roux2006geometric}. This concept can be introduced by referring to the Poincare sphere, that is a hypothetical sphere representing a variety of the polarisation states of light. We can attribute any change of polarisation of light to a trajectory on the sphere. If the polarisation state of light is changing in the way to follow a closed trajectory on the Poincare sphere and enclosing a solid angle $\Theta$, a light beam acquires an extra phase of $\pm\Theta/2$ where the sign depends on clockwise or counterclockwise motion along the closed trajectory. If the trajectory does not make a closed loop on the Poincare sphere, the accumulation of the geometric phase is found by connecting the starting and the final points with the shortest geodesic line~\cite{wang2017non}. As the possible total angle on the sphere does not exceed 4$\pi$, the maximal accumulation of the geometric
phase is 2$\pi$, that provides a full flexibility for the wavefront control.

{\bf Planar metadevices} based on the concepts of Huygens' surfaces, form birefringence, and geometric phase allows to achieve an arbitrary complex control of phase and polarisation of light in ultra-thin transparent nanostructured layers. Figures~4(a-f) show characteristic examples of functional flat metadevices. Recent demonstrations include beam deflectors~\cite{lin2014dielectric,yu2015high,arbabi2015dielectric,shalaev2015high,sell2017large,zhou2017efficient},
lenses~\cite{schonbrun2011reconfigurable,lin2014dielectric,vo2014sub,matsui2015flat,arbabi2015subwavelength,zhan2016,
khorasaninejad2016metalenses,arbabi2016miniature,arbabi2016multiwavelength,chen2017immersion,groever2017meta,
khorasaninejad2017achromatic,paniagua2017metalens,lalanne_review}
vortex masks~\cite{chong2015polarization,shalaev2015high,zhan2016,wang2017orbital} and
holograms~\cite{arbabi2015dielectric,wang2016grayscale,chong2016efficient,zhao2016dielectric,zhao2016full,
wang2016visible,khorasaninejad2016broadband,reviewer1}
that are based on the phase control with Huygens' sources and/or geometric phase. Holograms deserve a special attention as they rely on complex wavefront engineering and therefore showcase the potential of the meta-optics platform. An ability to control polarisation spun demonstration of waveplates~\cite{kruk2016invited}, polarization beam splitters~\cite{khorasaninejad2015efficient} and vector beam $q$-plates~\cite{arbabi2015dielectric,kruk2016invited}. Combination of phase and polarisation control leads to the development of novel types of optical elements, such as lenses with polarisation-switchable focus~\cite{schonbrun2011reconfigurable} and holograms with polarisation-dependant reconstructed image~\cite{mueller2017metasurface}.

\begin{figure*}
\includegraphics[width=0.9\textwidth]{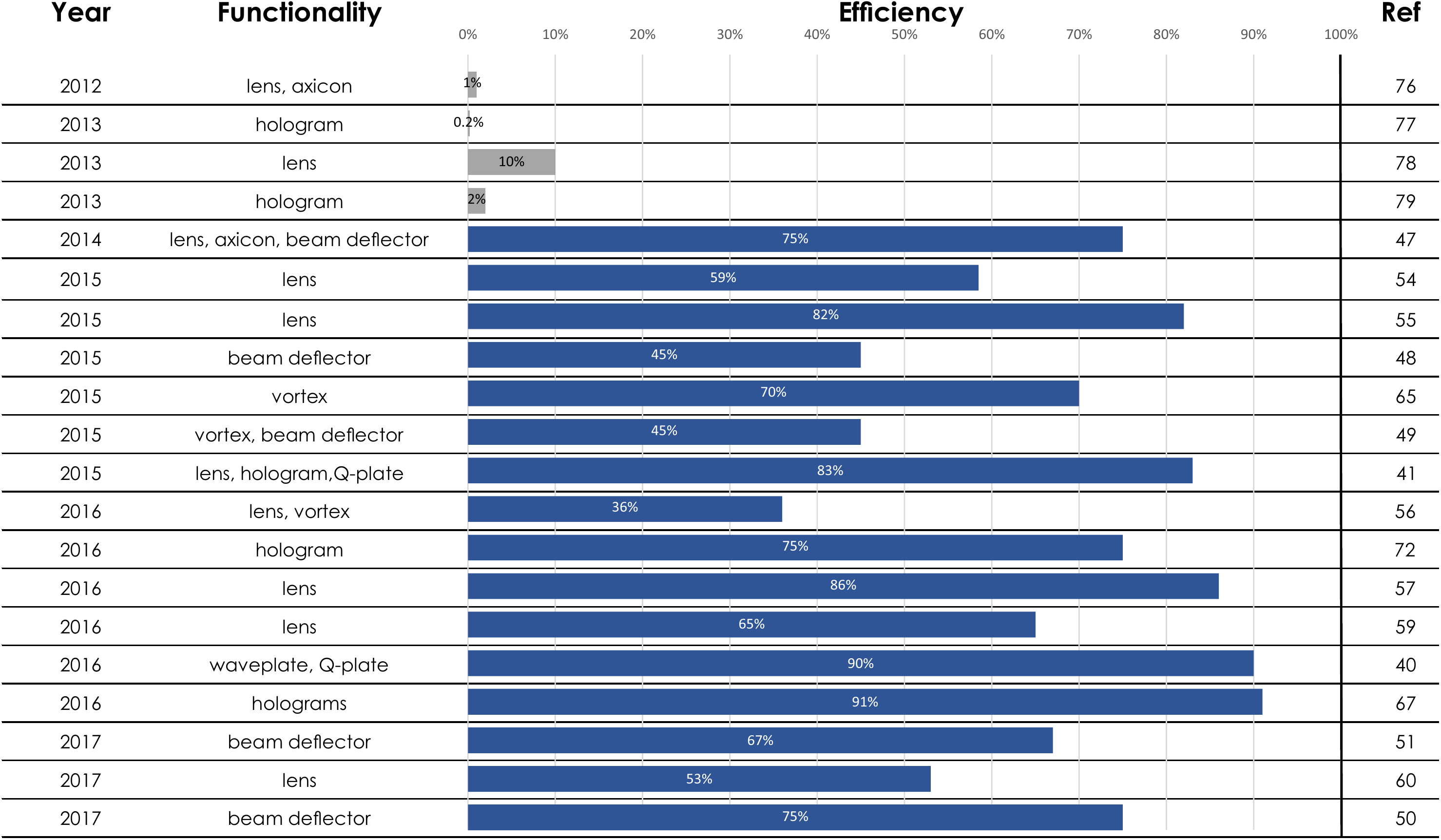}
\caption{\textbf{Efficiencies of transparent planar meta-optics devices.} Grey colour on the top marks the cases with elements made of metals,
whereas blue colour stands for dielectric planar metadevices. Last column provides the references to the corresponding papers.}
\label{fig:Part3Fig2}
\end{figure*}

Metasurfaces based on dielectric nanostructures with both electric and magnetic Mie-type resonances have resulted in the best efficiency to date for functional flat meta-optics. A diagram in Fig.~\ref{fig:Part3Fig2} visualises the rapid progress in planar meta-optics operating in {\em the transmission regime}.  Nowadays, all-dielectric metasurface have well-surpassed their plasmonic counterparts~\cite{plasmonicsreview,plasmonics1,plasmonics2,plasmonics3,plasmonics4,review1,review2,review3,review4,review5,review6}: dielectric metasurfaces with transparency exceeding 90$\%$~\cite{arbabi2015dielectric,kruk2016invited,wang2016grayscale} have been demonstrated, being {\em as transparent as a glass film}. Efficiencies of both phase and polarisation control employing generalized Huygens' principle have reached 99$\%$~\cite{kruk2016invited,wang2016grayscale}. Thus, in many cases, planar meta-optics matches or outperforms conventional optical elements offering much thinner elements.

Besides high performance, the metadevices provide many important advantages.
Meta-optics is ultra-thin. In contrast to conventional optical components that control light through its propagation inside bulk materials, the meta-optics controls the wave propagation via resonant response. That allows to achieve device thickness of only a fraction of a wavelength. That is more than a 100 times thinner than a human hair.

Polarization control with meta-optics provides effective birefringence that is several times larger than that in conventional crystals and liquid crystals. And spatial resolution of birefringence gradients is orders of magnitude higher than in liquid-crystal phase modulators let alone birefringent crystals.

Fabrication of highly-efficient planar meta-optics is easily  scalable as it requires only one patterning step that produces a structure with binary height. This is a big advantage especially when compared to conventional diffractive optics. For example, kinoform diffractive optical elements are structures of continuously varying height. Their fabrication requires grey-scale lithography techniques that impose challenges on reproducibility and scalability. Multi-level diffractive optics -- a "digitized" version of kinoforms -- is more reliable to fabricate, however it requires multiple lithography steps precisely aligned to reach high efficiencies. Exemplary, a multi-level diffractive optics components with a diffraction efficiency of 99$\%$ would require at least 16 levels of height, that is four lithography steps.

Another big advantage of metasurfaces compared to kinoforms and multi-level diffractive optics comes from their subwavelength thickness: metasurfaces are essentially flat for incident light, therefore the effects of shadowing do not occur.

\section*{Advanced metadevices}

Next, we look beyond passive and linear structures summarized above and foresee the great opportunities provided by all-dielectric resonant meta-optics in the rapidly developing fields of nonlinear nanophotonics, active meta-devices, tunable metasurfaces, quantum optics, and
the emerging field of topological photonics.
\begin{figure*}
\includegraphics[width=0.99\textwidth]{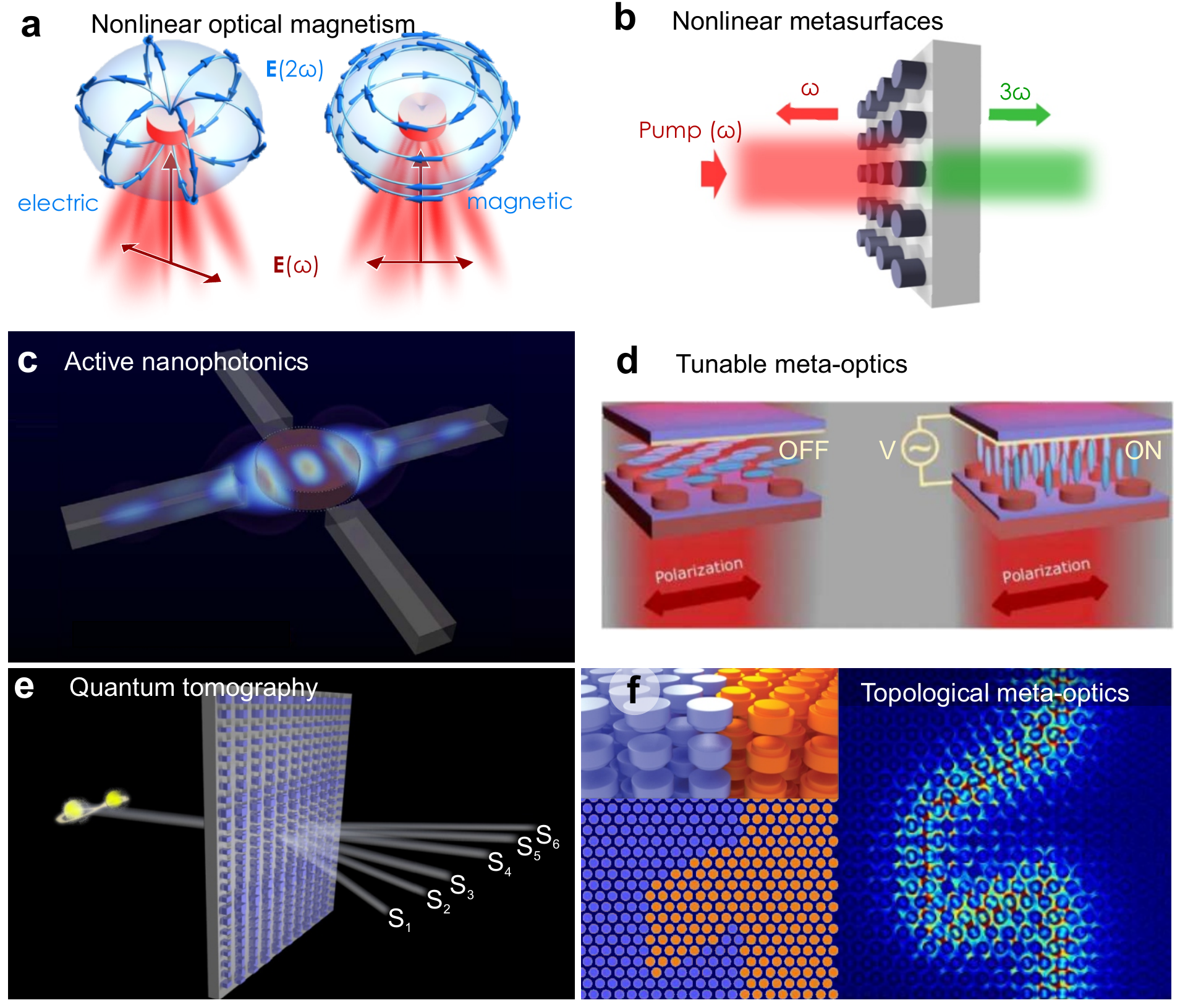}
\caption{\textbf{Advanced functionalities and metadevices.} (a) Nonlinear antennas generating electric or magnetic second harmonic~\cite{kruk2017nonlinear}.
(b) Nonlinear mirror metasurface with control over directionality of third harmonic and transmission/reflection of pump~\cite{wang2017nonlinear}.
(c) Nanolaser based on an anapole mode cavity~\cite{gongora2017anapole}.
(d) Electrically tunable metadevice based on dielectric metasurface infiltrated with liquid crystals~\cite{Komar}.
(e) Metasurface detector of quantum states of an arbitrary number of entangled photons~\cite{wang2017quantum}.
(f) Topological edge states in 3D resonant dielectric nanostructures~\cite{slobozhanyuk2016three}.
}
\label{fig:Fig_devices}
\end{figure*}

{\bf Nonlinear optics} describes intensity-dependent optical phenomena facilitated by strong light-matter interaction.
For many years, the nonlinear optics has been mostly studying electric interactions, while the magnetic side of electromagnetism was disregarded. This comes from negligible nonlinear magnetization of materials at a microscopic level. However, nowadays it's been recognised that an intrinsic microscopic nonlinear electric polarizability of resonant nanoparticles may lead to induced \emph{magnetic nonlinear effects}~\cite{smirnova2016multipolar}. Electric and magnetic multipolar generation can reshape completely nonlinear effects at the nanoscale. Recently nonlinear magnetic light emission was directly observed in second-harmonic generation from AlGaAs nanodisks [see Fig.~\ref{fig:Fig_devices}(a)].~\cite{kruk2017nonlinear} The presence of both electric and magnetic nonlinearities enrich the spectrum of interference effects and allow to enhance efficiency and control
directionality and polarisation of nonlinear processes~\cite{camacho2016nonlinear,wang2017shaping}. A natural extension of the field of multipolar nonlinear nanophotonics is the physics of nonlinear metasurfaces with novel functionalities. Recently a proposal has been made for a nonlinear mirror based~\cite{wang2017nonlinear} on THG in Si metasurfaces [see Fig.~\ref{fig:Fig_devices}(b)]. Nonlinear mirrors have been established as a universal tool for ultrafast control of laser dynamics. However, traditional approaches are based on bulk crystals posing major difficulties on phase matching and alignment. Ultra-thin nonlinear mirrors based on all-dielectric metasurfaces would offer a major breakthrough in nonlinear-optics technologies~\cite{review_nonlinear1,review_nonlinear2} with applications in nanoscale sensors, photodetectors, and light sources.

{\bf Active optical media} structured at the nanoscale bring new physics to lasing dynamics and interplays of gain and loss. Recently, the new concept of  a nanoscale laser has been suggested~\cite{gongora2017anapole} [see Fig.~\ref{fig:Fig_devices}(c)] based on a tightly confined anapole mode [see Fig. 3 (b) for the concept of anapole]. Designed as an optically pumped semiconductor nanodisk, it enables efficient coupling to waveguides and new mechanisms of mode-locking for ultrafast laser pulse generation, thus offering an attractive platform for advanced integrated photonic circuitry~\cite{gongora2017anapole}.

Importantly, the Mie resonances can govern the enhancement of spontaneous photon emission~\cite{bonod_2016} and photoluminescence~\cite{schilling_2017} from embedded quantum emitters such as quantum dots and NV centers, due to
a good spatial overlap with the excited Mie modes. This field is in its infant stage, but it should see a rapid progress.

{\bf Dynamically reconfigurable all-dielectric metadevices} open-up the whole new area of applications such as beam steering, holographic displays, optical zoom with flat lenses, and increase of information transfer and processing speed in telecommunication systems. The tunability can be achieved in several ways, and only some of them are mentioned below.

\emph{Liquid crystals} infiltration is the promising way to tune dielectric resonant nanoparticles as Mie resonances are sensitive to the refractive index of the surrounding. Therefore, the change of orientation, or phase of liquid crystals has a drastic effect on their optical responses. Both thermal~\cite{sautter2015active} and electric~\cite{minovich2012liquid,Komar} reconfigurability of meta-optics infiltrated with liquid crystals have been demonstrated [see Fig.~\ref{fig:Fig_devices}(d)].

\emph{Mechanical deformation} was suggested and demonstrated as a tunability mechanism in metasurfaces~\cite{gutruf2015mechanically,kamali2016highly}. Stretching or bending metasurfaces alters
inter-element coupling, and thus modifies the resulting optical response.

\emph{Free-charge carrier density} change can be used to modulate permittivity of some materials. This allows to tune amplitude, phase and polarisation of transmitted or/and reflected waves. Several material platforms have benn suggested, including highly-doped Si and Ge nanoparticles~\cite{lewi2015widely}, GaAs nanostructures~\cite{jun2012electrically} as well as graphene~\cite{samson2010metamaterial,reviewer2} and thin ITO layers~\cite{park2016dynamic,reviewer3}.

\emph{Phase changing materials} offer a large change of material parameters upon the structural transition between two states.  Typical examples include a transition from the amorphous to the crystalline state, or transition from insulator-to-conductor state.
Thus, reconfigurability of meta-optics was achieved by implementing different phase changing material such as germanium antimony telluride (GST)~\cite{michel2013using,wang2016optically,chu2016active}, indium antimonide (InSb)~\cite{michel2013using},  gallium lanthanum sulphide (GLS)~\cite{samson2010metamaterial}, and vanadium oxide  (VO$_2$)~\cite{driscoll2009memory,dicken2009frequency,zhu2017dynamically}.

\emph{Ultra-fast nonlinear all-optical switching} is a promising mechanism for modulation of metasurfaces' response. Strong nonlinear self-modulation was demonstrated for Si~\cite{shcherbakov2015ultrafast} and GaAs nanostructures~\cite{shcherbakov2017ultrafast}. In addition, recent demonstrations of fast and large nonlinearities of epsilon-near-zero materials such as ITO~\cite{alam2016large} and AZO~\cite{caspani2016enhanced} shed new light on all-optical tunability of meta-optics, and are expected to be developed further.

{\bf Quantum optics} is another promising direction for all-dielectric resonant nanophotonics. Recently, dielectric metasurfaces were suggested for multi-photon polarization tomography.~\cite{wang2017quantum} The control of quantum-polarization states of entangled photons is an essential part of quantum information technologies, but conventional methods of measurements of quantum photon states rely on reconfigurable optical elements in bulk setups that compromises the speed and miniaturization of the whole system. In contrast to that, a single passive ultra-thin metasurface can be used for the full reconstruction of pure or mixed quantum polarization states across a broad bandwidth [see Fig.~\ref{fig:Fig_devices}(e)] when paired with simple polarization-insensitive photon detectors. A subwavelength thin structure provides an ultimate miniaturization, and can facilitate photon tomography by spatially-resolved imaging without a need for reconfiguration. Such parallel-detection approach promises not only better robustness and scalability, but also the possibility to study the dynamics of quantum states in real time. We anticipate that metasurfaces will find applications in quantum communication, cryptography, and computation, where they can serve as robust and accurate elements, replacing multiple bulky optical components.

{\bf Topological nanophotonics} has emerged recently as a promising direction of modern optics that aims to explore the concepts of topological order for novel phenomena in light-matter interactions.~\cite{lu2014topological}. Topology studies global properties of systems that remain unchanged under continuous deformations of parameters, such as stretching or bending for the case of geometrical parameters. An interesting physics happens at an interface of two topologically dissimilar media: an edge state appears “bridging a gap” between two different topologies. Topological states in photonic systems correspond to localised or propagating modes of light. Having its topological origin, these modes are immune to deformations of an optical system. Topological photonics is seen as a promising direction to achieve insulation of optical circuitry from parasitic scattering on defects and interfaces. While first demonstrations of optical topological states were realised in systems of coupled optical waveguides~\cite{hafezi2013imaging}, resonant nanophotonics allows to create topological states on a subwavelength scale. To date, sub-wavelength edge states in chains of dielectric nanoparticles with Mie resonances have been suggested and demonstrated~\cite{slobozhanyuk2015subwavelength,kruk2017edge}. More recently, a proposal was made for all-dielectric meta-crystals consisting of resonant building blocks exhibiting electromagnetic duality between electric and magnetic fields that may be used to create photonic topological insulators supporting propagating edge states~\cite{slobozhanyuk2016three}. Figure~\ref{fig:Fig_devices}(f) shows a side view and a top view of an interface between two topologically different optical meta-crystals. And right side of the figure depicts light propagating along the interface with no scattering at sharp bending corners. In such systems the magneto-electrical coupling plays the role of a synthetic gauge field that determines a topological transition to an ‘insulating’ regime with a complete photonic band-gap. We envision the key role of loss-less resonant dielectric nanophotonics and metasurfaces in future demonstrations of topologically nontrivial phases of materials and edge states towards realisation of topologically protected waveguiding and routing.

\section*{Summary and outlook}

The recent progress in low-loss meta-optics and all-dielectric resonant nanophotonics is associated with the study of optical magnetism
in high-index dielectric  nanoparticles that emerged as a new direction in the modern photonics.  This field originates from century-old studies of light scattering by Gustav Mie and Lord Rayleigh, and it brings a powerful tool of optically-induced
electric and magnetic Mie resonances into the fields of nanophotonics and metamaterials with the functionalities enabled by optical magnetic
response.
All-dielectric resonant structures have many advantages over their plasmonic counterparts, including resonant behaviour and low energy dissipation into heat. As such, the designs and effects offered by high-index resonant nanophotonics are expected to complement or even substitute different plasmonic components of nanoscale optical structures and metasurfaces  for a range of applications.
Importantly, the existence of strong electric and magnetic dipole resonances and smart engineering of multipole resonances can result in constructive or destructive interferences with unusual beam shaping, and it may also lead to the resonant enhancement of magnetic fields in dielectric nanoparticles that bring many novel functionalities for both linear and nonlinear regimes.

We expect that all-dielectric resonant meta-optics will ultimately employ all of the advantages of optically-induced magnetic resonances and shape many important applications and functionalities, including optical sensing, parametric amplification, nonlinear active media, as well as will find applications in integrated quantum circuitry and topological photonics underpinning a new generation of nanoantennas, nanolasers, highly-efficient metasurfaces, and ultrafast metadevices.






{\bf Acknowledgements.} We thank our numerous colleagues and co-authors for many useful discussions and fruitful collaboration, the list is too long to name them all, but we appreciate their help and suggestions. This work was supported by different projects of the Australian Research Council and the ITMO University.

\end{document}